# Accelerate micromagnetic simulations with GPU programming in MATLAB

Ru Zhu[1]

Graceland University, Lamoni, Iowa 50140 USA

**Abstract**

A finite-difference Micromagnetic simulation code in MATLAB is presented with Graphics Processing Unit (GPU) acceleration. The high performance of Graphics Processing Unit (GPU) is demonstrated compared to a typical Central Processing Unit (CPU) based code. The speed-up of GPU to CPU is shown to be greater than 30 for problems with larger sizes on a mid-end GPU in single precision. The code is less than 200 lines and suitable for new algorithm developing.

## 1. Introduction

Micromagnetic simulations are indispensable tools to study magnetic dynamics and develop novel magnetic devices. Micromagnetic codes running on Central Processing Unit (CPU) such as OOMMF [1] and magpar [2] have been widely adopted in research of magnetism. Micromagnetic simulations of complex magnetic structures require fine geometrical discretization, and are time consuming.

To accelerate the simulation, several research groups have been working on applying general purpose Graphics Processing Units (GPU) programming in the fields of Micromagnetics [3–9]. Thanks to the high computing performance of GPU, great speed-up has been reported in these implementations.

However, most of these implementations are in-house codes and their source codes are not publicly available. The objective of this work is to implement a simple but complete micromagnetic code in MATLAB accelerated by GPU programming.

In this paper, section 2 lists the formulation of micromagnetics and discusses the implementation of MATLAB code, including the calculation of demagnetization field, exchange field and anisotropy field. Section 3 validates the simulation result with a micromagnetic standard problem and evaluates the speedup of this micromagnetic code at various problem sizes as compared with a popular CPU-based micromagnetic code.

[1] Email address: zhu@sting.graceland.edu.

## 2. Principle

To study magnetic dynamics, we need to define a magnetization vector $\vec{M} = (M_x, M_y, M_z)$ in a computational cell in the magnetic nanostructure. The saturation magnetization $M_s = \sqrt{M_x^2 + M_y^2 + M_z^2}$ in this computational cell is assumed to be constant. The magnetic energy density related to this vector can be written down as

$$\varepsilon = A[(\nabla \frac{M_x}{M_s})^2 + (\nabla \frac{M_x}{M_s})^2 + (\nabla \frac{M_x}{M_s})^2] + K_u \frac{(M_y^2 + M_z^2)}{M_s^2} - \frac{1}{2}(\mu_0 \vec{H}_{demag} \vec{M}) - (\mu_0 \vec{H}_{extern} \vec{M}) \quad (1)$$

which shows that the magnetic energy density has the following terms: the exchange, anisotropy, demagnetization and Zeeman energy densities, where $A$ is the material exchange constant, $K_u$ is the anisotropy constant, $\mu_0$ is the vacuum permeability, $H_{demag}$ is demagnetization field and $H_{extern}$ is the external field or Zeeman field. The anisotropy has an easy axis on the $x$ direction and is assumed to be uniaxial.

The dynamics of magnetization vector is governed by the Landau-Lifshitz-Gilbert (LLG) equation. In the low damping limit, it is:

$$\frac{d\vec{M}}{dt} = -\frac{\gamma}{1+\alpha^2}(\vec{M} \times \mu_0 H_{eff}) - \frac{\alpha\gamma}{(1+\alpha^2)M_s}[\vec{M} \times (\vec{M} \times \mu_0 \vec{H}_{eff})] \quad (2)$$

where $\alpha$ is the damping constant, $\gamma$ is the gyromagnetic ratio, and $H_{eff}$ is the effective magnetic field derived from the magnetic energy density:

$$\vec{H}_{eff} = -\frac{\delta\varepsilon}{\mu_0 \delta\vec{M}} = \vec{H}_{exch} + \vec{H}_{anis} + \vec{H}_{demag} + \vec{H}_{extern} \quad (3)$$

where $\frac{\delta\varepsilon}{\delta\vec{M}}$ is the functional derivative of $\varepsilon$ with respect to $\vec{M}$ . $\vec{H}_{exch}$ and $\vec{H}_{anis}$ are exchange field and anisotropy field respectively.

### 2.1 Calculation of Exchange Field

According to (1) and (3), the $x$ component of exchange field is

$$H_{exch,x} = \frac{2A}{\mu_0 M_s^2} \nabla^2 M_x . \quad (4)$$

In the calculation of exchange field $H_{exch}$, the computing region must be discretized carefully and the magnetizations of neighboring computational cells need to be taken into account. In this work, the entire computing region is divided into $n_x \times n_y \times n_z$ cells, each cell with an equal volume of $\delta \times \delta \times \delta$. The cells are labeled with indices

$1 \le i \le n_x$,

$1 \le j \le n_y$,

$1 \le k \le n_z$.

According to (4), the Cartesian components of exchange field can be calculated in a six-neighborhood scheme:

$$H_{exch,x} = \frac{2A}{\mu_0 M_s^{\ 2}} \{ \frac{M_x(i+1,j,k) - 2M_x(i,j,k) + M_x(i-1,j,k)}{\delta^2} + \frac{M_x(i,j+1,k) - 2M_x(i,j,k) + M_x(i,j-1,k)}{\delta^2} + \frac{M_x(i,j,k+1) - 2M_x(i,j,k) + M_x(i,j,k-1)}{\delta^2} \}. \tag{5}$$

Other components of $\vec{H}_{exch}$ can be obtained by replacing $x$ with $y$ or $z$ in (5).

The calculation of exchange field on the boundary is handled with Neumann boundary condition

$$\frac{\partial M_x}{\partial \vec{n}} = \frac{\partial M_y}{\partial \vec{n}} = \frac{\partial M_z}{\partial \vec{n}} = 0 \tag{6}$$

In other words, the non-existing neighbor cell on the boundary will be replaced by the cell at the center. For example, if we calculate exchange field on the left boundary where $i = 1$, then $M_x(0,j,k) = M_x(1,j,k)$, and exchange field is calculated as

$$\begin{aligned} H_{exch,x}(i=1,j,k) &= \frac{2A}{\mu_0 M_s^{\ 2}} \{ \frac{M_x(2,j,k) - 2M_x(1,j,k) + M_x(1,j,k)}{\delta^2} \\ &+ \frac{M_x(i,j+1,k) - 2M_x(i,j,k) + M_x(i,j-1,k)}{\delta^2} \\ &+ \frac{M_x(i,j,k+1) - 2M_x(i,j,k) + M_x(i,j,k-1)}{\delta^2} \} \\ &= \frac{2A}{\mu_0 M_s^{\ 2}} \{ \frac{M_x(2,j,k) - M_x(1,j,k)}{\delta^2} \\ &+ \frac{M_x(i,j+1,k) - 2M_x(i,j,k) + M_x(i,j-1,k)}{\delta^2} \\ &+ \frac{M_x(i,j,k+1) - 2M_x(i,j,k) + M_x(i,j,k-1)}{\delta^2} \}. \end{aligned} \tag{7}$$

Assuming that $j \neq 1, j \neq n_y$, and $k \neq 1, k \neq n_z$. The MATLAB code is given in listing 1. Note that we calculate exchange field by creating duplicates of magnetization matrix instead of doing it in place. This sacrifices some memory space but speeds up the calculation on GPU, where simple algorithm is preferred on its parallel hardware architecture.

**Listing 1.** Calculation of exchange field. `Hx0,1,2,3,4,5` are helper matrices that hold the magnetizations of neighboring cells in a six-neighborhood scheme.

```
Hx0 (2:end,:,:) = Mx(1:end-1,:,:); % left neighbor
Hx0 (1,:,:) = Hx0(2,:,:); % Neumann boundary condition
Hx1 (1:end-1,:,:) = Mx(2:end,:,:); % right neighbor
Hx1 (end,:,:) = Hx1(end-1,:,:);
Hx2 (:,2:end,:) = Mx(:,1:end-1,:); % back neighbor
Hx2 (:,1,:) = Hx2(:,2,:);
Hx3 (:,1:end-1,:) = Mx(:,2:end,:); % front neighbor
Hx3 (:,end,:) = Hx3(:,end-1,:);
Hx4 (:,:,2:end) = Mx(:,:,1:end-1); % bottom neighbor
Hx4 (:,:,1) = Hx4(:,:,2);
Hx5 (:,:,1:end-1) = Mx(:,:,2:end); % top neighbor
Hx5 (:,:,end) = Hx5(:,:,end-1);
exch = 2 * exchConstant / mu_0 / Ms / Ms; % dd is the cell size
Hx = Hx + exch / dd / dd * (Hx0 + Hx1 + Hx2 + Hx3 + Hx4 + Hx5 - 6 * Mx);
```

### 2.2 Calculation of Anisotropy Field

According to (1) and (3),

$$H_{anis,x} = \frac{2K_u}{\mu_0 M_s^2} M_x. \tag{8}$$

Assuming $H_K = \frac{2K_u}{\mu_0 M_s}$, then $H_{anis,x} = \frac{H_K}{M_s} M_x$. Other components of the anisotropy field can be obtained by permutation of the index. The MATLAB code is shown in listing 2.

**Listing 2.** Calculation of anisotropy field.

```
Hx_anis = 100; % in kA/m

Hx_anis = gpuArray(Hx_anis); % copy data from CPU to GPU

Hx = Hx + Hx_anis / Ms * Mx;
```

### 2.3 Calculation of Demagnetization Field

Calculation of demagnetization field is the most time-demanding part of micromagnetic simulation. A time complexity of $O(N^2)$ is needed for brute force evaluation of a micromagnetic sample with N computational cells. It may be reduced to $O(N \log N)$ with fast Fourier transform (FFT) [10]. This work calculates demagnetization field with the FFT algorithm and the detail is discussed below.

Demagnetization field is caused by the dipole interaction of magnetization in the sample studied. Its formulation has been studied by several research groups [11, 12]. The formula given by [11] is adopted in this work:

$$\begin{pmatrix} H_{demag,x} \\ H_{demag,y} \\ H_{demag,z} \end{pmatrix} = \Sigma \begin{pmatrix} K_{xx} & K_{xy} & K_{xz} \\ K_{yx} & K_{yy} & K_{yz} \\ K_{zx} & K_{zy} & K_{zz} \end{pmatrix} \begin{pmatrix} M_x \\ M_y \\ M_z \end{pmatrix} \tag{9}$$

where $K$ is demagnetization tensor and can be calculated as

$$K_{xx}(I,J,K) = \frac{1}{4\pi} \sum_{i=1}^{1} \sum_{j=0}^{1} \sum_{k=0}^{1} (-1)^{i+j+k} \tan^{-1} \frac{(K+k-\frac{1}{2})(J+j-\frac{1}{2})\delta}{r_{ijk}(I+i-\frac{1}{2})} \tag{10}$$

$$K_{xy}(I,J,K) = K_{yx}(I,J,K) = -\frac{1}{4\pi} \sum_{i=1}^{1} \sum_{j=0}^{1} \sum_{k=0}^{1} (-1)^{i+j+k} \log\left[(K+k-\frac{1}{2})\delta + r_{ijk}\right] \tag{11}$$

Other components can be calculated by permutations of indices.

**Listing 3.** Calculation of Demagnetization tensor.

```
prefactor = 1 / 4 / 3.14159265;
for K = -nz + 1 : nz - 1 % Calculation of Demag tensor
    for J = -ny + 1 : ny - 1
        for I = -nx + 1 : nx - 1
            if I == 0 && J == 0 && K == 0
                continue
            end
            L = I + nx; % shift the indices, b/c no negative index allowed in
MATLAB
            M = J + ny;
            N = K + nz;
            for i = 0 : 1 % helper indices
                for j = 0 : 1
                    for k = 0 : 1
                        r = sqrt ( (I+i-0.5)*(I+i-0.5)*dd*dd +(J+j-0.5)*(J+j-
0.5)*dd*dd +(K+k-0.5)*(K+k-0.5)*dd*dd);
                        Kxx(L,M,N) = Kxx(L,M,N) + (-1).^(i+j+k) * atan( (K+k-
0.5) * (J+j-0.5) * dd / r / (I+i-0.5));
```

```
                        Kxy(L,M,N) = Kxy(L,M,N) + (-1).^(i+j+k) * log( (K+k-0.5)
* dd + r);
                        Kxz(L,M,N) = Kxz(L,M,N) + (-1).^(i+j+k) * log( (J+j-0.5)
* dd + r);
                        Kyy(L,M,N) = Kyy(L,M,N) + (-1).^(i+j+k) * atan( (I+i-
0.5) * (K+k-0.5) * dd / r / (J+j-0.5));
                        Kyz(L,M,N) = Kyz(L,M,N) + (-1).^(i+j+k) * log( (I+i-0.5)
* dd + r);
                        Kzz(L,M,N) = Kzz(L,M,N) + (-1).^(i+j+k) * atan( (J+j-
0.5) * (I+i-0.5) * dd / r / (K+k-0.5));
                    end
                end
            end
            Kxx(L,M,N) = Kxx(L,M,N) * prefactor;
            Kxy(L,M,N) = Kxy(L,M,N) * - prefactor;
            Kxz(L,M,N) = Kxz(L,M,N) * - prefactor;
            Kyy(L,M,N) = Kyy(L,M,N) * prefactor;
            Kyz(L,M,N) = Kyz(L,M,N) * - prefactor;
            Kzz(L,M,N) = Kzz(L,M,N) * prefactor;
        end
    end
end % calculation of demag tensor done
```

As mentioned before, the sample studied is divided into $n_x \times n_y \times n_z$ cells. In this case, demagnetization field can be calculated as

$$H_{demag}(i,j,k) = \sum_{l=1}^{n_x}\sum_{m=1}^{n_y}\sum_{n=1}^{n_z} M(l,m,n)K(l-i,m-j,n-k) \tag{12}$$

or more specific,

$$\begin{aligned} H_{demag,x}(i,j,k) &= \sum_{l=1}^{n_x}\sum_{m=1}^{n_y}\sum_{n=1}^{n_z}\{M_x(l,m,n)K_{xx}(l-i,m-j,n-k) \\ &+ M_y(l,m,n)K_{xy}(l-i,m-j,n-k) \\ &+ M_z(l,m,n)K_{xz}(l-i,m-j,n-k)\} \end{aligned} \tag{13}$$

Other components of $H_{demag}$ can be obtained by replacing x with y or z in (13).

According to (13) demagnetization field matrix is a convolution of magnetization matrix and demagnetization tensor. By applying DFT theorem to both sides of the equation, we can get

$$\begin{aligned} \tilde{H}_{demag,x}(i,j,k) &= \tilde{M}_x(i,j,k)\cdot\tilde{K}_{xx}(i,j,k) \\ &+ \tilde{M}_y(i,j,k)\cdot\tilde{K}_{xy}(i,j,k) \\ &+ \tilde{M}_z(i,j,k)\cdot\tilde{K}_{xz}(i,j,k) \end{aligned} \tag{14}$$

Finally, demagnetization field $H_{demag}$ can be obtained by taking the inverse FFT of $\tilde{H}_{demag}$.

For non-periodic boundary condition, zero padding of magnetization and demagnetization data required to avoid circular convolution. After zero padding the input data size increases to $2n_x \times 2n_y \times 2n_z$, as demonstrated by Fig. 1, and the MATLAB code is shown in listing 4, where `fftn` and `ifftn` are discrete Fourier transforms and inverse transform of multi-dimensional data.

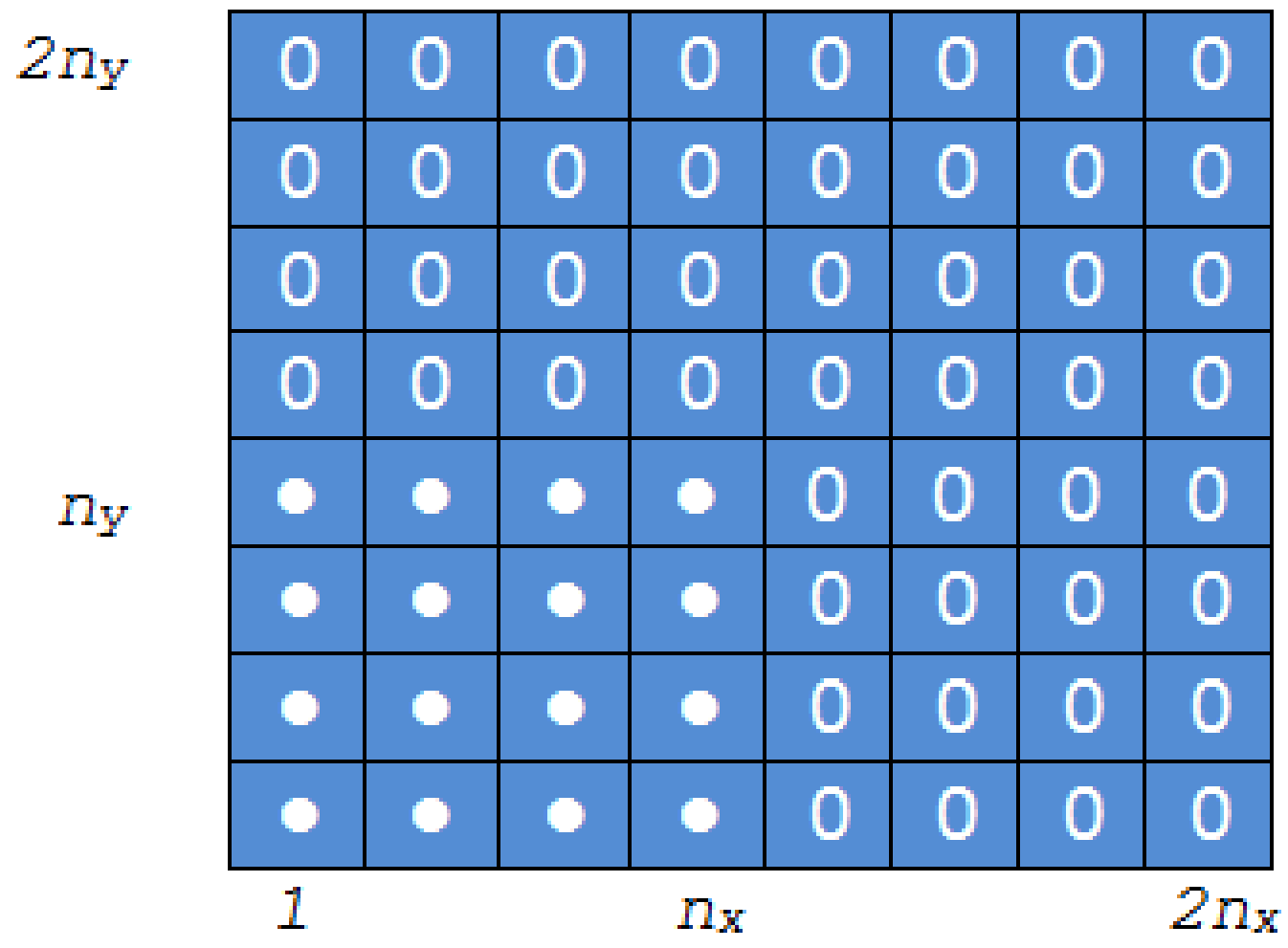

Fig. 1 A cross-sectional view of a rectangular magnetic sample after zero-padding.

**Listing 4.** Zero padding of magnetization data and calculation of demagnetization field.

```
Mx(end + nx, end + ny, end + nz) = 0; % zero padding
My(end + nx, end + ny, end + nz) = 0;
Mz(end + nx, end + ny, end + nz) = 0;
Hx = ifftn(fftn(Mx) .* Kxx_fft + fftn(My) .* Kxy_fft + fftn(Mz) .* Kxz_fft);
Hy = ifftn(fftn(Mx) .* Kxy_fft + fftn(My) .* Kyy_fft + fftn(Mz) .* Kyz_fft);
Hz = ifftn(fftn(Mx) .* Kxz_fft + fftn(My) .* Kyz_fft + fftn(Mz) .* Kzz_fft);
```

### 2.4 Acceleration by GPU programming

However, for large problem sizes, the simulation time can be intolerably long due to limited computing power of CPU. Therefore to boost the performance of micromagnetic simulation, the computing power of GPU can be utilized.

The hardware architecture of GPUs is intrinsically different from CPUs. Due to its large number of Arithmetic Logic Units (ALU), GPU can significantly increase the performance of simulation on parallel algorithms. Since the FFT algorithm can be implemented as parallel, it is suitable to apply GPU programming to micromagnetic simulation. The comparison between CPU and GPU is schematically illustrated by Fig. 2.

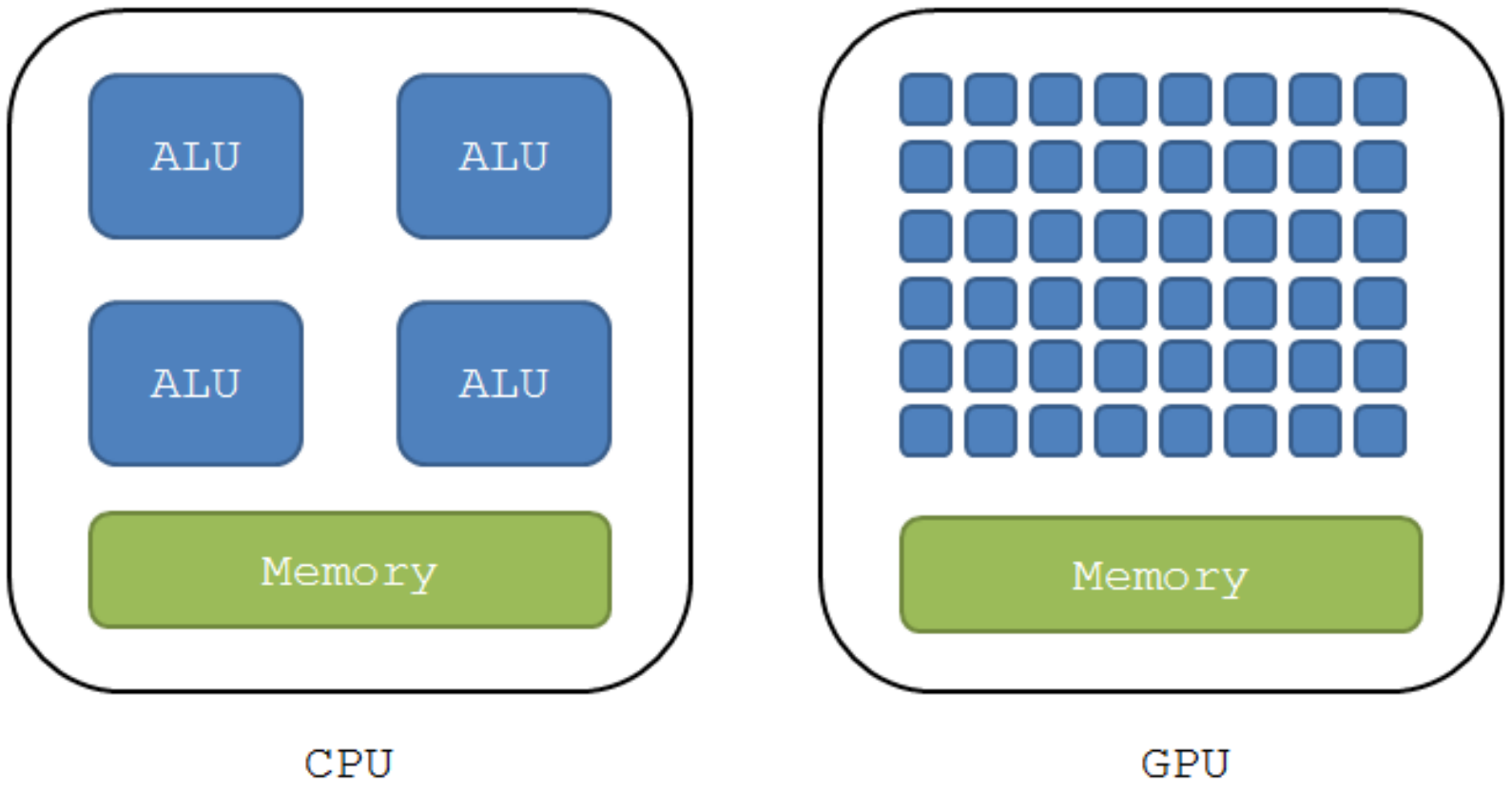

Fig. 2 A comparison between Hardware architectures of CPU and GPU. There are more ALUs dedicated to data processing on GPU.

The bottleneck of GPU computing is usually the data transfer between CPU and GPU. The data I/O between GPU and CPU is about 10 GB/s, as compared to the data I/O is between ALUs of GPU and its own memory (> 100 GB/s). To fully utilize the computing power of GPU, in the micromagnetic code presented, the initial condition of the sample is set on the GPU, and all calculation in done on GPU. Data is only transferred back to CPU periodically for output purpose. Listing 5 presents the initialization of magnetization on GPU.

**Listing 5.** Magnetization data set up on GPU.

```
Mx = repmat(Ms, [nx ny nz]); % magnetization on x direction
Mx = gpuArray(Mx);
My = gpuArray.zeros(nx, ny, nz);
Mz = My;
```

### 3. Results and Discussion

To validate the simulation result, the μmag standard problem #4 [13], field 1 was used. It was proposed by the μMag modeling group and assumed a sample with a size of 500nm × 125nm × 3nm. The exchange constant of the sample $A = 1.3 \times 10^{-11} J/m$ , saturation magnetization of $M_S = 8.0 \times 10^5 A/m$ and there is no anisotropy. The sample is relaxed to S-state by setting a saturating field along the (1, 1, 1) direction and then slowly reduce it. Then an external field of (-24.6 mT, 4.3 mT, 0 mT) is applied to reverse the magnetization. The code used in this validation can be found in Append. A. Figure 3 shows the close agreement of simulation result and the result of OOMMF.

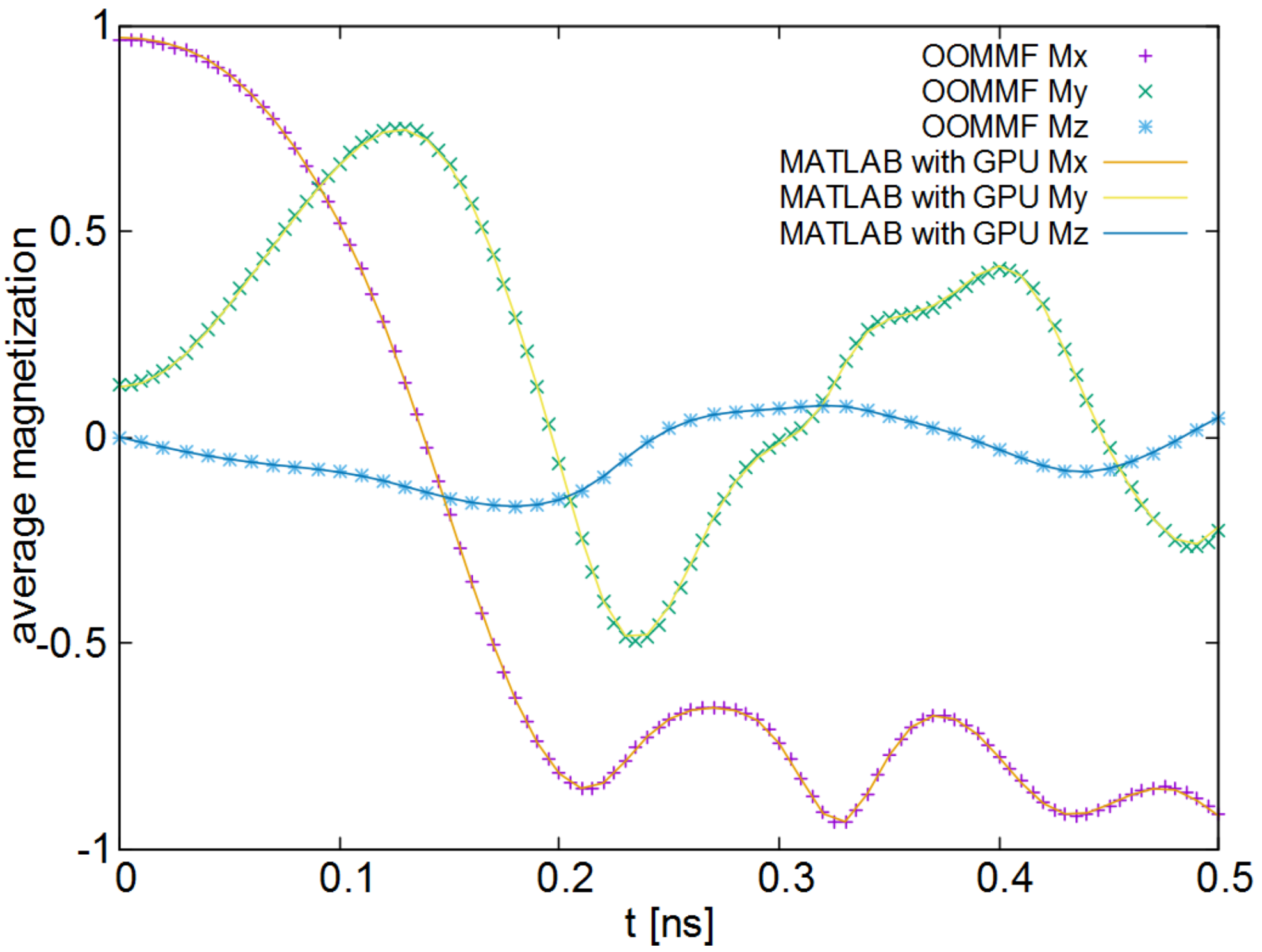

**Fig. 3 Comparison of the simulation results of OOMMF and this work for μmag standard problem #4, field 1. The horizontal axis is simulation time in nanoseconds, and the vertical axis is components of average magnetization.**

The micromagnetic standard problem #3 [14] was used to evaluate the speedup of simulation gained by GPU in this code. In this problem, a cubic magnetic sample with exchange constant $A = 1\times10^{-11} J/m$, saturation magnetization $M_s = 1000 kA/m$ and anisotropy field $H_{anis} = 100 kA/m$ is discretized to $N\times N\times N$ cells and the minimum energy state is reached by applying the LLG equation with a large damping constant. Demagnetization field, exchange field and uniaxial anisotropy field all affect the time evolution of magnetization in the sample.

To benchmark the code, an NVidia GTX 650 Ti running on Intel Xeon E5410 CPU with 4GB of RAM was used. The GPU is a middle-end product which costs less than $100 on the consumer market. For comparison, the benchmark of CPU version of this code on the same hardware platform is also presented. While this code can solve problems of any size limited by the graphic memory allocable by the GPU, dimensions with powers of two were benchmarked to examine the performance of code varying with problem size, as shown in table 1 and figure 4-5.

**Table 1**. Per-step simulation time needed by CPU and GPU solvers for different 3D problem sizes with the Euler algorithm. Numbers are in milliseconds.

| Size | CPU | GPU double precision | Speedup | GPU single precision | Speedup |
|---|---|---|---|---|---|
| $8^3$ | 3.99 | 24.15 | ×0.14 | 22.58 | ×0.14 |
| $16^3$ | 26.56 | 26.97 | ×1.1 | 23.97 | ×1.2 |
| $32^3$ | 248.8 | 45.21 | ×5.5 | 26.89 | ×10 |
| $64^3$ | 2903 | 238 | ×12 | 76.24 | ×38 |

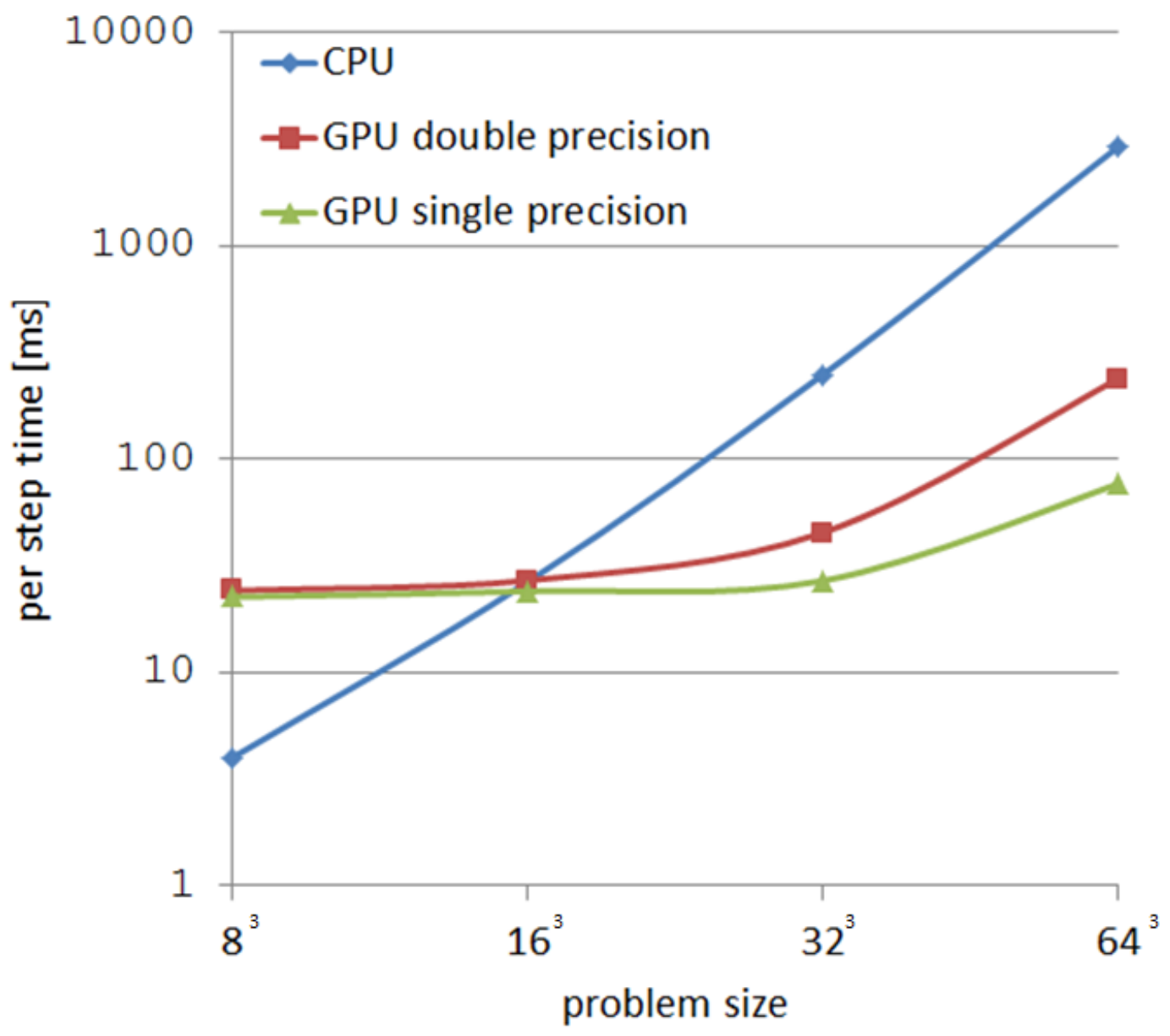

**Fig. 4 Per-step simulation time at various 3D problem sizes. The μmag standard problem #3 was used in this benchmark.**

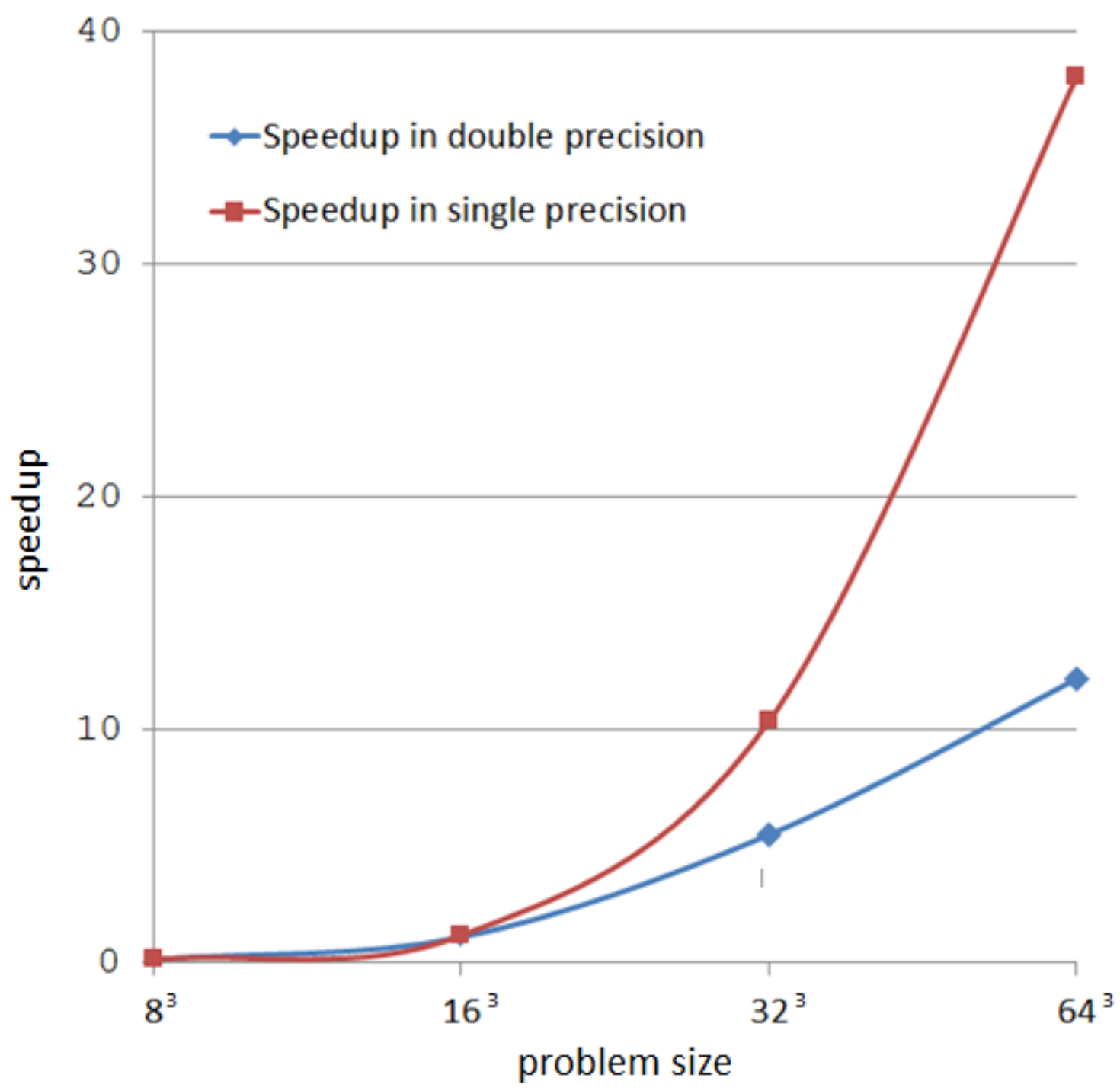

**Fig. 5 Speedup of GPU code running on NVidia GTX 650 Ti as compared to CPU code running on Intel Xeon E5410.**

It is noticeable that at smaller problem sizes the GPU code performance is not much better or even worse than CPU solver. This can be explained by two factors. The first factor is the kernel launching overhead. This overhead is constant regardless of the problem size, so it becomes predominant when the problem size decreases. The second factor is parallel processing architecture of GPU. The NVidia GTX 650 Ti has 768 steam processors, and they are mostly idle at a small problem size, leaving the computing power of GPU not fully utilized.

It can also be observed that performance of the GPU code in single precision is significantly better than that in double precision. This is because GPUs such as GTX 650 Ti is designed to meet the requirement of consumer market which emphasizes on its graphics processing capability, which is done predominantly in single precision. The author expects to see a boost in double precision performance when the code runs on professional GPUs such as NVidia TITAN series.

## 4. Summary

A GPU-accelerated micromagnetic code is presented to address the slow speed problem of large simulation problems. The speed boost relative to CPU simulations is significant at problem with large input sizes. This code is short but complete and can serve as foundation of future algorithm development. The full micromagnetic code is available from [15] and also given in Append. A.

## Acknowledgements

This work is supported by Graceland University professional development program.

**References**

[1] Donahue, Michael Joseph, and Donald Gene Porter. OOMMF User's guide. US Department of Commerce, Technology Administration, National Institute of Standards and Technology, 1999.

[2] Scholz, Werner, et al. "Scalable parallel micromagnetic solvers for magnetic nanostructures." Computational Materials Science 28.2 (2003): 366-383.

[3] Kakay, Attila, Elmar Westphal, and Riccardo Hertel. "Speedup of FEM micromagnetic simulations with Graphics Processing Units." Magnetics, IEEE Transactions on 46.6 (2010): 2303-2306.

[4] Vansteenkiste, Arne, and Ben Van de Wiele. "MuMax: a new high-performance micromagnetic simulation tool." Journal of Magnetism and Magnetic Materials 323.21 (2011): 2585-2591.

[5] Chang, R., et al. "FastMag: Fast micromagnetic solver for complex magnetic structures." Journal of Applied Physics 109.7 (2011): 07D358.

[6] Li, Shaojing, Boris Livshitz, and Vitaliy Lomakin. "Graphics processing unit accelerated micromagnetic solver." Magnetics, IEEE Transactions on 46.6 (2010): 2373-2375.

[7] Lopez-Diaz, L., et al. "Micromagnetic simulations using graphics processing units." Journal of Physics D: Applied Physics 45.32 (2012): 323001.

[8] Zhu, Ru. "Speedup of Micromagnetic Simulations with C++ AMP On Graphics Processing Units." arXiv preprint arXiv:1406.7459 (2014).

[9] Zhu, Ru. "Grace: a Cross-platform Micromagnetic Simulator On Graphics Processing Units." arXiv preprint arXiv:1411.2565 (2014).

[10] Hayashi, Nobuo, Koji Saito, and Yoshinobu Nakatani. "Calculation of demagnetizing field distribution based on fast Fourier transform of convolution." Japanese journal of applied physics 35.12A (1996): 6065-6073.

[11] Nakatani, Yoshinobu, Yasutaro Uesaka, and Nobuo Hayashi. "Direct solution of the Landau-Lifshitz-Gilbert equation for micromagnetics." Japanese Journal of Applied Physics 28.12R (1989): 2485.

[12] Newell, Andrew J., Wyn Williams, and David J. Dunlop. "A generalization of the demagnetizing tensor for nonuniform magnetization." Journal of Geophysical Research: Solid Earth (1978–2012) 98.B6 (1993): 9551-9555.

[13] McMichael, R. D., et al. "Switching dynamics and critical behavior of standard problem No. 4." Journal of Applied Physics 89.11 (2001): 7603-7605.

[14] Michael Donahue, et al. "µMAG Standard Problem #3." µMAG organization. Mar 1998. Web. 23 Jun. 2014

[15] "A GPU accelerated micromagnetic simulator in MATLAB." `https://github.com/cygnusc/mumag.matlab`

**Append. A.** GPU accelerated simulation code for standard problem #4, field 1.

---

```
nx = 166; % number of cells on x direction
ny = 42;
nz = 1;
dd = 3; % cell volume = dd x dd x dd
dt = 5E-6; % timestep in nanoseconds
dt = gpuArray(dt); % copy data from CPU to GPU
timesteps = 150000;
alpha = 0.5; % damping constant to relax system to S-state
alpha = gpuArray(alpha);
exchConstant = 1.3E-11 * 1E18; % nanometer/nanosecond units

mu_0 = 1.256636; % vacuum permeability, = 4 * pi / 10
Ms = 800; % saturation magnetization
Ms = gpuArray(Ms);
exch = 2 * exchConstant / mu_0 / Ms / Ms;
exch = gpuArray(exch);
prefactor1 = (-0.221) * dt / (1 + alpha * alpha);
prefactor2 = prefactor1 * alpha / Ms;
prefactor1 = gpuArray(prefactor1);
prefactor2 = gpuArray(prefactor2);

Mx = repmat(Ms, [nx ny nz]); % magnetization on x direction
Mx = gpuArray(Mx);
My = gpuArray.zeros([nx ny nz]);
Mz = My;

deltaMx = gpuArray.zeros([nx ny nz]);
deltaMy = gpuArray.zeros([nx ny nz]);
deltaMz = gpuArray.zeros([nx ny nz]);
mag = gpuArray.zeros([nx ny nz]);

Kxx = zeros(nx * 2, ny * 2, nz * 2); % Initialization of demagnetization tensor
Kxy = Kxx;
Kxz = Kxx;
Kyy = Kxx;
Kyz = Kxx;
Kzz = Kxx;
prefactor = 1 / 4 / 3.14159265;
for K = -nz + 1 : nz - 1 % Calculation of Demag tensor
    for J = -ny + 1 : ny - 1
        for I = -nx + 1 : nx - 1
            if I == 0 && J == 0 && K == 0
                continue
            end
            L = I + nx; % shift the indices, b/c no negative index allowed in
MATLAB
            M = J + ny;
            N = K + nz;
            for i = 0 : 1 % helper indices
                for j = 0 : 1
                    for k = 0 : 1
```

```
                        r = sqrt ( (I+i-0.5)*(I+i-0.5)*dd*dd +(J+j-0.5)*(J+j-
0.5)*dd*dd +(K+k-0.5)*(K+k-0.5)*dd*dd);
                        Kxx(L,M,N) = Kxx(L,M,N) + (-1).^(i+j+k) * atan( (K+k-
0.5) * (J+j-0.5) * dd / r / (I+i-0.5));
                        Kxy(L,M,N) = Kxy(L,M,N) + (-1).^(i+j+k) * log( (K+k-0.5)
* dd + r);
                        Kxz(L,M,N) = Kxz(L,M,N) + (-1).^(i+j+k) * log( (J+j-0.5)
* dd + r);
                        Kyy(L,M,N) = Kyy(L,M,N) + (-1).^(i+j+k) * atan( (I+i-
0.5) * (K+k-0.5) * dd / r / (J+j-0.5));
                        Kyz(L,M,N) = Kyz(L,M,N) + (-1).^(i+j+k) * log( (I+i-0.5)
* dd + r);
                        Kzz(L,M,N) = Kzz(L,M,N) + (-1).^(i+j+k) * atan( (J+j-
0.5) * (I+i-0.5) * dd / r / (K+k-0.5));
                    end
                end
            end
            Kxx(L,M,N) = Kxx(L,M,N) * prefactor;
            Kxy(L,M,N) = Kxy(L,M,N) * - prefactor;
            Kxz(L,M,N) = Kxz(L,M,N) * - prefactor;
            Kyy(L,M,N) = Kyy(L,M,N) * prefactor;
            Kyz(L,M,N) = Kyz(L,M,N) * - prefactor;
            Kzz(L,M,N) = Kzz(L,M,N) * prefactor;
        end
    end
end % calculation of demag tensor done

Kxx_fft = fftn(Kxx); % fast fourier transform of demag tensor
Kxy_fft = fftn(Kxy); % need to be done only one time
Kxz_fft = fftn(Kxz);
Kyy_fft = fftn(Kyy);
Kyz_fft = fftn(Kyz);
Kzz_fft = fftn(Kzz);

Kxx_fft = gpuArray(Kxx_fft);
Kxy_fft = gpuArray(Kxy_fft);
Kxz_fft = gpuArray(Kxz_fft);
Kyy_fft = gpuArray(Kyy_fft);
Kyz_fft = gpuArray(Kyz_fft);
Kzz_fft = gpuArray(Kzz_fft);

Hx_exch = gpuArray.zeros(nx, ny, nz);
Hy_exch = gpuArray.zeros(nx, ny, nz);
Hz_exch = gpuArray.zeros(nx, ny, nz);

outFile = fopen('Mdata.txt', 'w');

Hx0 = gpuArray.zeros(nx, ny, nz);
Hx1 = gpuArray.zeros(nx, ny, nz);
Hx2 = gpuArray.zeros(nx, ny, nz);
Hx3 = gpuArray.zeros(nx, ny, nz);

Hy0 = gpuArray.zeros(nx, ny, nz);
Hy1 = gpuArray.zeros(nx, ny, nz);
Hy2 = gpuArray.zeros(nx, ny, nz);
```

```
Hy3 = gpuArray.zeros(nx, ny, nz);

Hz0 = gpuArray.zeros(nx, ny, nz);
Hz1 = gpuArray.zeros(nx, ny, nz);
Hz2 = gpuArray.zeros(nx, ny, nz);
Hz3 = gpuArray.zeros(nx, ny, nz);

for t = 1 : timesteps
    Mx(end + nx, end + ny, end + nz) = 0; % zero padding
    My(end + nx, end + ny, end + nz) = 0;
    Mz(end + nx, end + ny, end + nz) = 0;

    Hx = ifftn(fftn(Mx) .* Kxx_fft + fftn(My) .* Kxy_fft + fftn(Mz) .*
Kxz_fft); % calc demag field with fft
    Hy = ifftn(fftn(Mx) .* Kxy_fft + fftn(My) .* Kyy_fft + fftn(Mz) .* Kyz_fft);
    Hz = ifftn(fftn(Mx) .* Kxz_fft + fftn(My) .* Kyz_fft + fftn(Mz) .* Kzz_fft);

    Hx = real(Hx (nx:(2 * nx - 1), ny:(2 * ny - 1), nz:(2 * nz - 1) ) ); %
truncation of demag field
    Hy = real(Hy (nx:(2 * nx - 1), ny:(2 * ny - 1), nz:(2 * nz - 1) ) );
    Hz = real(Hz (nx:(2 * nx - 1), ny:(2 * ny - 1), nz:(2 * nz - 1) ) );
    Mx = Mx (1:nx, 1:ny, 1:nz); % truncation of Mx, remove zero padding
    My = My (1:nx, 1:ny, 1:nz);
    Mz = Mz (1:nx, 1:ny, 1:nz);
    % calculation of exchange field
    Hx0 (2:end,:,:) = Mx(1:end-1,:,:);
    Hx0 (1,:,:) = Hx0(2,:,:);
    Hx1 (1:end-1,:,:) = Mx(2:end,:,:);
    Hx1 (end,:,:) = Hx1(end-1,:,:);

    Hx2 (:,2:end,:) = Mx(:,1:end-1,:);
    Hx2 (:,1,:) = Hx2(:,2,:);
    Hx3 (:,1:end-1,:) = Mx(:,2:end,:);
    Hx3 (:,end,:) = Hx3(:,end-1,:);

    Hy0 (2:end,:,:) = My(1:end-1,:,:);
    Hy0 (1,:,:) = Hy0(2,:,:);
    Hy1 (1:end-1,:,:) = My(2:end,:,:);
    Hy1 (end,:,:) = Hy1(end-1,:,:);

    Hy2 (:,2:end,:) = My(:,1:end-1,:);
    Hy2 (:,1,:) = Hy2(:,2,:);
    Hy3 (:,1:end-1,:) = My(:,2:end,:);
    Hy3 (:,end,:) = Hy3(:,end-1,:);

    Hz0 (2:end,:,:) = Mz(1:end-1,:,:);
    Hz0 (1,:,:) = Hz0(2,:,:);
    Hz1 (1:end-1,:,:) = Mz(2:end,:,:);
    Hz1 (end,:,:) = Hz1(end-1,:,:);

    Hz2 (:,2:end,:) = Mz(:,1:end-1,:);
    Hz2 (:,1,:) = Hz2(:,2,:);
    Hz3 (:,1:end-1,:) = Mz(:,2:end,:);
    Hz3 (:,end,:) = Hz3(:,end-1,:);
```

```
    Hx = Hx + exch / dd / dd * (Hx0 + Hx1 + Hx2 + Hx3 - 4 * Mx);
    Hy = Hy + exch / dd / dd * (Hy0 + Hy1 + Hy2 + Hy3 - 4 * My);
    Hz = Hz + exch / dd / dd * (Hz0 + Hz1 + Hz2 + Hz3 - 4 * Mz);

    if t < 4000
        Hx = Hx + 100; % apply a saturation field to get S-state
        Hy = Hy + 100;
        Hz = Hz + 100;
    elseif t < 6000
        Hx = Hx + (6000 - t) / 20; % gradually diminish the field
        Hx = Hx + (6000 - t) / 20;
        Hx = Hx + (6000 - t) / 20;
    elseif t > 50000
        Hx = Hx - 19.576; % apply the reverse field
        Hy = Hy + 3.422;
        alpha = 0.02;
        prefactor1 = (-0.221) * dt / (1 + alpha * alpha);
        prefactor2 = prefactor1 * alpha / Ms;
    end
    % apply LLG equation
    MxHx = My .* Hz - Mz .* Hy; % = M cross H
    MxHy = Mz .* Hx - Mx .* Hz;
    MxHz = Mx .* Hy - My .* Hx;

    deltaMx = prefactor1 .* MxHx + prefactor2 .* (My .* MxHz - Mz .* MxHy);
    deltaMy = prefactor1 .* MxHy + prefactor2 .* (Mz .* MxHx - Mx .* MxHz);
    deltaMz = prefactor1 .* MxHz + prefactor2 .* (Mx .* MxHy - My .* MxHx);

    Mx = Mx + deltaMx;
    My = My + deltaMy;
    Mz = Mz + deltaMz;

    mag = sqrt(Mx .* Mx + My .* My + Mz .* Mz);
    Mx = Mx ./ mag * Ms;
    My = My ./ mag * Ms;
    Mz = Mz ./ mag * Ms;

    if mod(t, 1000) == 0 % recod the average magnetization
        MxMean = mean(Mx(:));
        MyMean = mean(My(:));
        MzMean = mean(Mz(:));
        fprintf(outFile, '%d\t%f\t%f\t%f\r\n', t, MxMean/Ms, MyMean/Ms,
MzMean/Ms);
    end
end
fclose('all');
```